\documentclass[a4paper,fleqn]{cas-dc}
\usepackage[english]{babel}
\usepackage[numbers]{natbib}
\usepackage{subfig}
\usepackage{xfrac}
\usepackage{siunitx}
\DeclareSIUnit{\rev}{rev}

\def\tsc#1{\csdef{#1}{\textsc{\lowercase{#1}}\xspace}}
\tsc{WGM}
\tsc{QE}

\begin{document}
\let\WriteBookmarks\relax
\def\floatpagepagefraction{1}
\def\textpagefraction{.001}

\shorttitle{Tool wear effects on turned surfaces via ISO 21920-2}

\shortauthors{Müller et al.}

\title [mode = title]{Effects of Tool Wear on the Surface Texture in Turning: A Feature Characterization Approach Based on ISO 21920-2}

\tnotemark[1] 
\tnotetext[1]{This work was funded by the Deutsche Forschungsgemeinschaft (DFG, German Research Foundation) - project numbers 461839204 and 521380776 (research priority program SPP 2402).}

\author[1]{Alexander Müller}[
    type=editor,
    style=german,
    orcid=0009-0004-5099-6296]
    \credit{Conceptualization, Data curation, Formal analysis, Investigation, Methodology, Software, Validation, Visualization, Writing -- original draft}
    \ead{alexander.mueller@mv.rptu.de}
    \cormark[1]
\author[2]{Maximilian Berndt}[
    orcid=0009-0006-2593-0302]
    \credit{Data curation, Investigation, Methodology, Visualization, Writing -- original draft}
\author[1]{Hagen Schmidt}[
    orcid=0009-0002-9868-4027]
    \credit{Methodology, Writing -- original draft}
\author[4]{Lars Müller}
    \credit{Writing -- review \& editing}
\author[3]{Matthias Eifler}[
    orcid=0000-0001-6628-7284]
    \credit{Writing -- review \& editing}
\author[4]{Eberhard Kerscher}[
    orcid=0000-0001-7211-1306]
    \credit{Funding acquisition, Supervision, Writing -- review \& editing}
\author[2]{Benjamin Kirsch}[
    orcid=0000-0001-7602-2511]
    \credit{Funding acquisition, Supervision, Writing -- review \& editing}
\author[1]{Jörg Seewig}[
    orcid=0000-0002-1420-1597]
    \credit{Funding acquisition, Supervision, Writing -- review \& editing}



\affiliation[1]{organization={Institute for Measurement and Sensor Technology, RPTU University Kaiserslautern-Landau},
    addressline={Gottlieb-Daimler-Str}, 
    postcode={67663}, 
    city={Kaiserslautern},
    country={Germany}}
\affiliation[2]{organization={Institute for Manufacturing Technology and Production Systems, RPTU University Kaiserslautern-Landau},
    addressline={Gottlieb-Daimler-Str}, 
    postcode={67663}, 
    city={Kaiserslautern},
    country={Germany}}
\affiliation[3]{organization={IU University of Applied Sciences},
    addressline={Juri-Gagarin-Ring 152}, 
    postcode={99084}, 
    city={Erfurt},
    country={Germany}}
\affiliation[4]{organization={Materials Testing, RPTU University Kaiserslautern-Landau},
    addressline={Gottlieb-Daimler-Str}, 
    postcode={67663},
    city={Kaiserslautern},
    country={Germany}}

\cortext[1]{Corresponding author}



\begin{abstract}
The surface texture of a turned component is a direct reflection of the manufacturing process and, in particular, of the tool cutting edge, acting like a fingerprint of both the process parameters and the tool wear condition. However, reading this fingerprint is only possible to a limited extent using classical surface parameters such as $R_\mathrm{a}$ or $R_\mathrm{q}$, as these describe the topography globally and do not allow for a spatially resolved evaluation of the process-induced deterministic structures. This work investigates the extent to which the feature characterization standardized in ISO 21920-2 can make this wear information accessible and physically interpretable, providing a basis for inline measurement strategies. The database consists of roughness profiles of twelve AlTiN-coated tungsten carbide indexable inserts (CNMG120408) machining normalized AISI 1045 steel, measured across nine wear states throughout the entire tool life, with three replicate profiles per state.
The correlation of standardized field and feature parameters with measures of crater wear, flank wear, and cutting time is first examined. Watershed segmentation is then adapted to extract the rotational tool grooves and statistically evaluate their geometry. Building on this, a newly developed mean-feature approach decomposes the profile into a deterministic and a stochastic component.
The analysis shows that wear-induced changes are almost entirely carried by the deterministic component, and within it by the trailing flank of the cutting groove. A direct comparison with confocal measurements confirms that the mean feature reconstructs the engaged section of the cutting edge geometry, with the trailing-flank steepening attributable to notch wear on the secondary cutting edge. An exhaustive evaluation of more than \num{920000} feature characterization combinations and multivariate models reveals that the groove-level mean maximum absolute gradient $\overline{R_\mathrm{dt}}_\mathrm{groove}$ alone already explains 83--90\,\% of the variance of the measured wear indicators, so that a single, physically motivated parameter is sufficient for robust wear estimation. Building on this, a follow-up study will investigate inline monitoring of tool wear using scattered light sensors.
\end{abstract}

\begin{highlights}
\item ISO 21920-2 feature characterization links surface profile to cutting edge geometry
\item Mean-feature approach separates deterministic and stochastic profile components
\item Wear-induced changes localize on the trailing flank of the cutting groove
\item Confocal measurements confirm notch wear as the dominant wear mechanism
\item Groove-level $\overline{R_\mathrm{dt}}_\mathrm{groove}$ enables robust wear estimation (83--90\,\% variance)
\end{highlights}



\maketitle

\section{Introduction} \label{introduction}

Tool wear is a critical factor affecting process stability, workpiece quality, and cost-effectiveness in machining operations such as turning \cite{Pusavec2010}. As wear progresses, the tribological conditions at the tool cutting edge change, potentially degrading surface quality and increasing process forces. Accurate quantification and characterization of tool wear state is therefore essential to optimize tool life and ensure high component quality.

In the idealized limit case, the profile of a turned surface is a geometric image of the tool cutting edge advancing along the feed direction, determined by the feed rate and the geometry of the engaged cutting edge \cite{Klocke2018, Sung2018}. From a kinematic perspective, the profile is a periodic repetition of the cutting edge geometry.

However, a measured profile fundamentally does not correspond to this purely kinematic image. In real processes, ploughed and elastic spring back effects can occur, so the cutting edge is only partially imaged ~\cite{Klocke2018}. Additionally, plastic lateral material flow, built-up edge formation, and tool-workpiece vibrations introduce further, largely stochastic deviations~\cite{Klocke2018}. Concurrently, progressive tool wear continuously reshapes the engaged cutting edge. The real profile is thus a superposition of a deterministic, process-kinematic component and a stochastic component, from which any wear-induced changes must be extracted.

Numerous experimental studies have investigated how this initial condition changes with increasing tool wear. Research on various workpiece materials, coated and uncoated indexable inserts, and varying cutting conditions consistently reports that conventional roughness parameters change measurably over tool life, yet disagreement exists regarding the direction and magnitude of these changes \cite{Grzesik2008, Liang2019, Das2017, Saini2012}. Height parameters such as $R_\mathrm{a}$, $R_\mathrm{q}$, and $R_\mathrm{z}$ are most commonly examined. Some authors report monotonic increases with progressing wear \cite{Kuntoglu2020}, while others observe less pronounced, non-monotonic behavior \cite{Saini2012, Derani2021}. The sensitivity of these parameters appears to depend strongly on the dominant wear mechanism, workpiece material, and specific cutting conditions, limiting their transferability across different experimental setups \cite{Khamel2012, Lima2005, Kumar2003, Bhushan2020}.

Since the cutting edge is imaged in the profile, wear-induced changes are in principle accessible from the machined surface, and several studies infer tool wear state directly from measured workpiece profiles \cite{Shahabi2008, Lim2022}. Conventional field parameters such as $R_\mathrm{a}$ or $R_\mathrm{q}$, however, compress the entire profile into a single global value, averaging over all measurement points. They therefore allow no spatially resolved evaluation of individual geometrical features such as single feed grooves, so that locally confined wear effects are lost in the averaging and the geometric wear information remains aggregated and inaccessible. This may explain the contradictory empirical findings reported above.
\begin{figure}
    \centering
    \includegraphics[width=0.95\columnwidth]{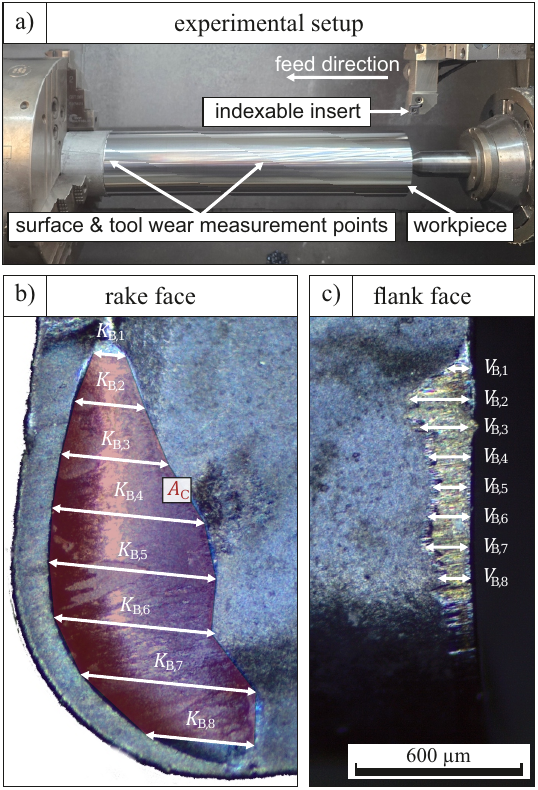}
    \caption{Experimental setup and wear quantification: (a) machining setup, (b) measurement of crater wear area $A_\mathrm{C}$ on the rake face, (c) measurements for the mean width of flank wear land $\overline{V}_\mathrm{B}$ on the flank face.}
    \label{fig:machining_setup}
\end{figure}
\begin{figure}
    \centering
    \includegraphics[width=0.95\columnwidth]{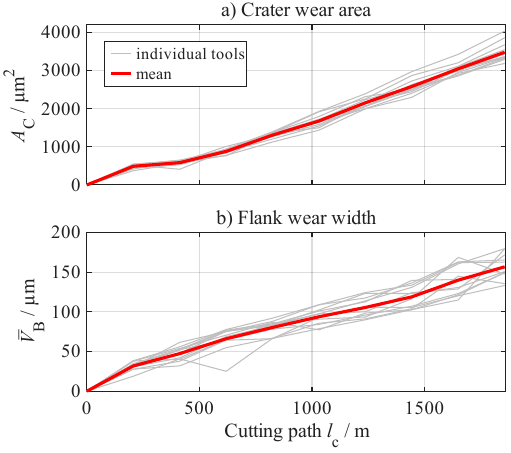}
    \caption{Progression of the crater wear area $A_\mathrm{C}$ (a) and the mean width of flank wear land $\overline{V}_\mathrm{B}$ (b) over the cutting path $l_\mathrm{c}$ for all twelve tools. Grey lines denote individual tools, while the red line marks the mean across all tools.}
    \label{fig:wear_progression}
\end{figure}
In contrast, feature parameters can be applied to individual profile segments, enabling position-dependent characterization of features that potentially correlate with tool wear. In 2021, ISO 21920-2 \cite{21920-2} introduced a feature characterization method in which characteristic features are extracted and quantified from the profile using watershed segmentation. An implementation guideline for this method forms the methodological foundation of this work \cite{Müller2025}. A spatially resolved characterization that specifically evaluates the groove geometry and relates observed changes to defined regions of the cutting edge has so far been largely absent for the turning process. The present work addresses this gap with an explicitly diagnostic objective.

To this end, the standardized profile feature characterization of ISO 21920-2 is transferred to turning and evaluated at the level of the individual feed groove. A mean-feature decomposition first separates the deterministic groove geometry from the stochastic profile component. Confocal measurements of the cutting edge then validate that the reconstructed groove shape images the engaged cutting edge and that the dominant wear-induced change, a steepening of the trailing flank, originates from notch wear on the secondary cutting edge. Finally, a single physically motivated parameter, the groove-level mean maximum absolute gradient $\overline{R_\mathrm{dt}}_\mathrm{groove}$, is identified and benchmarked against an exhaustive search over more than \num{920000} feature-characterization combinations and against multivariate models, and is shown to be sufficient for robust wear estimation. The samples analyzed here originate from earlier tool-life experiments by the authors \cite{Berndt2025}, in which a grey-box approach predicted tool wear from acoustic emission and cutting-force signals \cite{Schmidt2026}. The present work instead exploits the machined surface as a direct wear indicator. The insights gained provide a spatially resolved database to validate and enhance that grey-box model and establish the physical basis for new in-process strategies for the online determination and prediction of tool wear.

The structure of this work is as follows. Section~\ref{sec:method} describes the experimental setup, wear quantification, and topography measurements. Section~\ref{sec:default_parameters} correlates the standard field and feature parameters of ISO 21920-2 with the recorded wear measures. Section~\ref{sec:tool_grooves} adapts watershed segmentation to extract tool grooves. Building on this, Section~\ref{sec:mean_feature} introduces the mean-feature approach, which decomposes each profile into deterministic and stochastic components, and Section~\ref{sec:linking} links the resulting mean feature to the engaged cutting edge geometry via confocal measurements. Section~\ref{sec:additional} extends the analysis through systematic evaluation of feature characterization combinations and multivariate prediction models, before Section~\ref{sec:conclusion} summarizes the results and provides an outlook on future work.

\section{Experimental methodology} \label{sec:method}
The following section briefly summarizes the setup of the underlying tool life experiments \cite{Berndt2025} for completeness.

\subsection{Machining Setup and Process Parameters}
The turning experiments were conducted on a CNC turning machine (Boehringer NG200) under dry cutting conditions. Normalized AISI 1045 steel workpieces with a diameter of \SI{78}{\milli\metre} and a length of \SI{400}{\milli\metre} were used. The chemical composition and hardness of the workpiece material are provided in \cite{Berndt2025}. Longitudinal turning was performed at a cutting speed of $v_\mathrm{c}=\SI{250}{\metre/\minute}$, a feed rate of $f=\SI{0.2}{\milli\metre/\rev}$ (corresponding to a feed velocity of $v_\mathrm{f}\approx\SI{212}{\milli\metre/\minute}$), and a depth of cut of $a_\mathrm{p}=\SI{1.5}{\milli\metre}$, with a feed path of \SI{350}{\milli\metre} per workpiece. The machining setup, including the workpiece, indexable insert, and tool holder mounted in the machine, is shown in Fig.~\ref{fig:machining_setup}(a).

Aluminum titanium nitride (AlTiN) coated tungsten carbide indexable inserts of type CNMG120408RP (Kennametal) were used as cutting tools. The inserts feature an orthogonal rake angle $\gamma=\ang{-6}$, inclination angle $\lambda=\ang{-6}$, clearance angle $\alpha=\ang{6}$, approach angle $\kappa_r=\ang{95}$, and wedge angle $\beta=\ang{90}$. The tool life experiments were repeated twelve times. The end of tool life was reached when the manufactured surface showed significant quality deterioration, the coating exhibited large-scale failure, or the cutting edge chipped, as described in \cite{Berndt2025}.

\begin{figure}
    \centering
    \includegraphics[width=5cm]{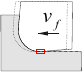}
    \caption{Schematic representation of the contact between the indexable insert and the workpiece, showing the cutting edge area relevant to the surface profile (region of interest).}
    \label{fig:indexable_insert_in_contact}
\end{figure}
\begin{figure}
    \centering
    \includegraphics{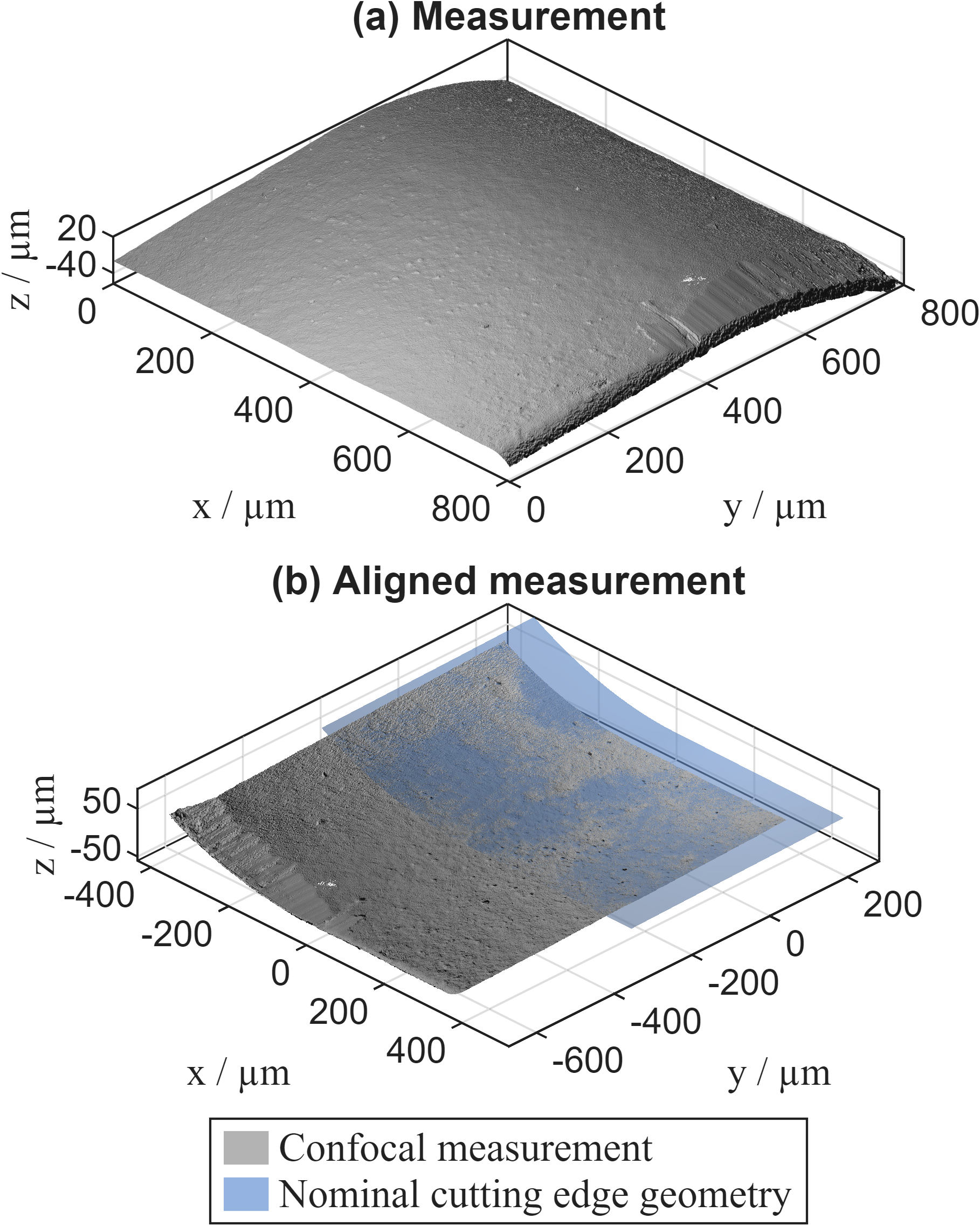}
    \caption{Alignment of the confocal cutting edge measurement to the workpiece coordinate system: a) raw confocal measurement of the indexable insert, b) ideal nominal geometry of the measured area, and c) aligned measurement after registration to the nominal geometry using the unworn region as reference.}
    \label{fig:confocal_measurement}
\end{figure}
\begin{figure*}
    \centering
    \includegraphics[scale=0.95]{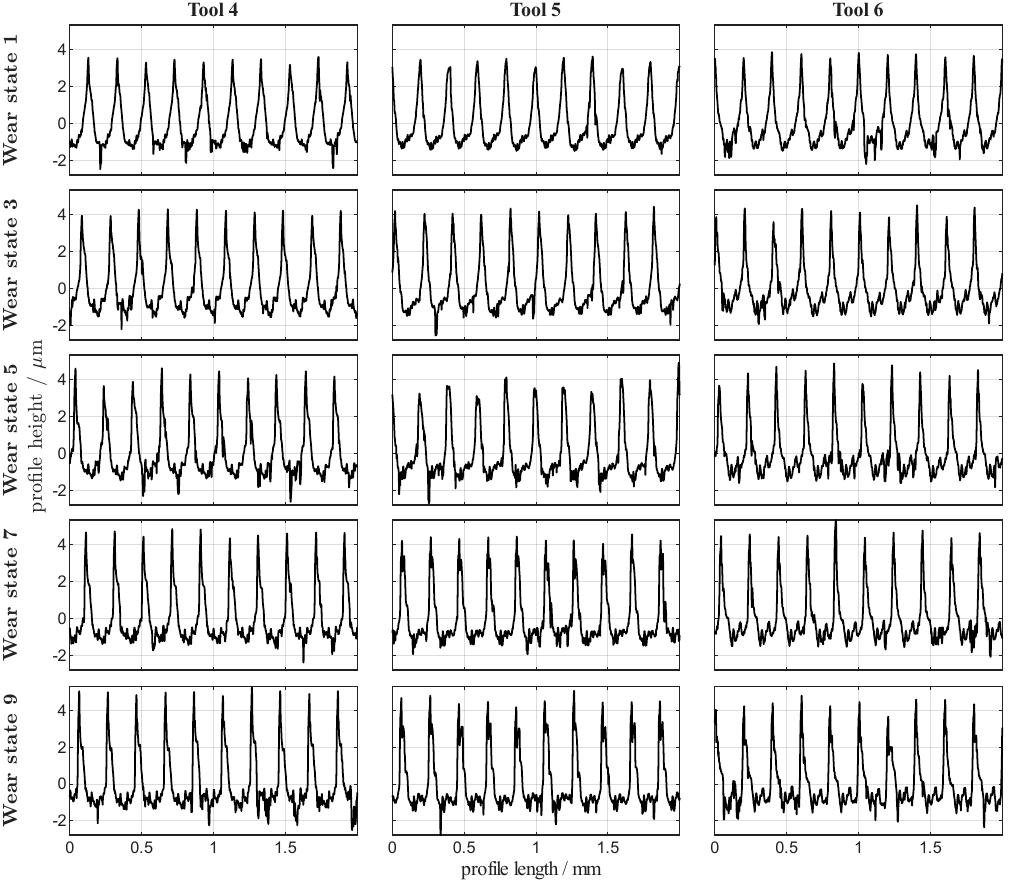}
    \caption{Roughness profiles of every second wear state of three different turning tools.}
    \label{fig:profiles}
\end{figure*}

\subsection{Wear Quantification}
Tool wear was quantified using two complementary indicators, measured at nine wear states equally spaced over the tool life, i.e.\ every \SI{206.2}{\metre} of cutting path. The mean width of flank wear land $\overline{V}_\mathrm{B}$ was determined as the arithmetic mean of eight individual flank wear measurements distributed uniformly across the active cutting edge within the depth of cut $a_\mathrm{p}$, as illustrated in Fig.~\ref{fig:machining_setup}(c). Measurements were taken using an Olympus SZ61 optical microscope at $3.5\times$ magnification. The crater wear area $A_\mathrm{C}$ characterizes the wear area on the rake face and is approximated from eight crater width measurements distributed equidistantly perpendicular to the cutting edge within $a_\mathrm{p}$, as illustrated in Fig.~\ref{fig:machining_setup}(b). The resulting seven partial areas are summed using a trapezoidal approximation to yield $A_\mathrm{C}$, as described in detail in \cite{Berndt2025}. Figure~\ref{fig:wear_progression} shows the resulting progression of both indicators over the cutting path for all twelve tools. Both increase steadily and consistently across the tool population, with $\overline{V}_\mathrm{B}$ following a largely linear trend and $A_\mathrm{C}$ a slightly progressive one.

\subsection{Topography measurements}
In addition to the tool wear measurements, three surface profiles per wear state were measured on the workpiece simultaneously, as depicted in Fig.~\ref{fig:machining_setup}(a).
The resulting 324 raw surface profiles were measured using a Mahr MarSurf XR20 GD120 profilometer with a diamond stylus of \SI{2}{\micro\metre} tip radius and \ang{90} cone angle (probe arm BFW A 4-45-2/90) over a sampling length of \SI{5.6}{\milli\metre}. In accordance with ISO 21920-2 \cite{21920-2} and ISO 21920-3 \cite{21920-3}, using setting class Sc3, the profiles were filtered as follows:
\begin{enumerate}
 \item S-filter: A linear Gaussian filter was applied according to ISO 16610-21 \cite{16610-21}, using a nesting index $N_\mathrm{is}=\SI{2.5}{\micro\metre}$ to remove small lateral-scale components.
 \item Form removal (F-operator): The nominal form was removed using a total least squares straight line fit, ensuring that form deviations do not influence the roughness evaluation.
 \item L-filter: A second linear Gaussian filter (ISO 16610-21 \cite{16610-21}) was used with a nesting index $N_\mathrm{ic}=\SI{0.8}{\milli\metre}$ to remove large lateral-scale (long-wavelength) waviness components.
\end{enumerate}
Figure~\ref{fig:profiles} shows representative roughness profiles for three different tools. For clarity, only every second wear state is depicted. Additionally, only half of each profile is plotted to improve the visibility of the surface structure.

\subsection{Cutting edge measurement}
The cutting edge region relevant for the surface profile 
(see Fig.~\ref{fig:indexable_insert_in_contact}) 
was measured using the NanoFocus µSurf confocal microscope ($20\times$ objective, $\mathrm{NA} = 0.6$). The resulting measurement area is $0.8 \times 0.8$\,\si{\milli\metre} with $1200 \times 1200$ height values. To extract the profile actually in contact with the workpiece, the measurement was transformed into the workpiece coordinate system, using the unworn region of the flank face as a reference for registration to the target geometry (Fig.~\ref{fig:confocal_measurement}).

\begin{figure}
    \centering
    \includegraphics{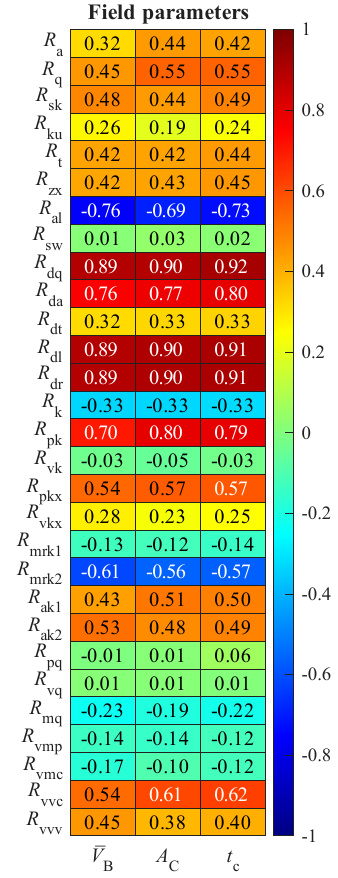}
    \caption{Pearson correlation of default field parameters. See Table~\ref{tab:iso_parameters} for parameter definitions.}
    \label{fig:corr_field}
\end{figure}
\begin{figure}
    \centering
    \includegraphics{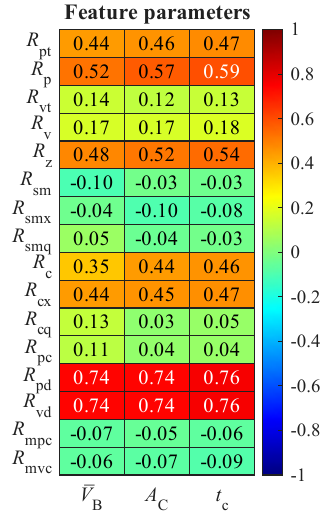}
    \caption{Pearson correlation of default feature parameters. See Table~\ref{tab:iso_parameters} for parameter definitions.}
    \label{fig:corr_feature}
\end{figure}

\begin{figure}[pos=p]
    \centering
    \includegraphics[scale=0.9,page=1]{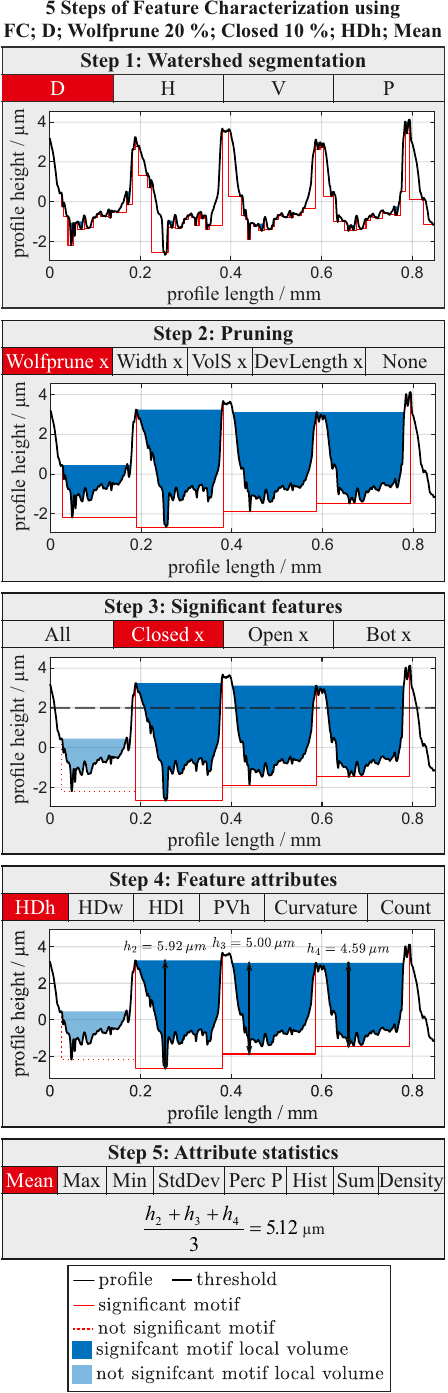}
    \caption{Five steps of feature characterization.}
    \label{fig:steps}
\end{figure}

\section{Correlation of default parameters of ISO 21920-2 with wear indicators}\label{sec:default_parameters}
Initially, all surface texture parameters defined in ISO 21920-2 \cite{21920-2} (see Table~\ref{tab:iso_parameters} in Appendix~\ref{appendix:figures} for an overview of the parameters and their definitions) were computed for the measured surface profiles using the default evaluation settings specified in ISO 21920-3 \cite{21920-3}. The correlation between these parameters and key wear indicators was then examined statistically: the crater wear area ($A_\mathrm{C}$), mean width of flank wear land ($\overline{V}_\mathrm{B}$), and cutting time ($t_\mathrm{c}$). To quantify the strength of linear relationships, the Pearson correlation coefficient $r$ was calculated for each pairing. These coefficients pool all 324 profiles and thus ignore the cluster structure induced by the twelve tools and the three replicate measurements per state. They serve here as purely descriptive screening measures for wear sensitivity, whereas the predictive performance of the parameters is validated tool-wise in Section~\ref{sec:multivariate}. Unlike $A_\mathrm{C}$ and $\overline{V}_\mathrm{B}$, which are measured directly and therefore carry measurement scatter, $t_\mathrm{c}$ increases monotonically in equal steps by construction and thus represents a noise-free, monotone substitute for wear progression. Its consistently highest correlation coefficients should accordingly be interpreted as an upper bound, whereas the directly measured indicators $A_\mathrm{C}$ and $\overline{V}_\mathrm{B}$ provide the more meaningful test of a parameter's sensitivity to wear.

It should be noted that the parameters derived from the material ratio (Abbott--Firestone) curve---in particular the $R_\mathrm{k}$ family ($R_\mathrm{k}$, $R_\mathrm{pk}$, $R_\mathrm{vk}$) and the related material-probability and volume parameters---presuppose a stratified, plateau-like surface with an S-shaped material ratio curve, as found on plateau-honed surfaces. For the periodic feed-groove profiles studied here the material ratio curve is close to linear, so these parameters lack a physical basis and their correlations are not interpreted.

Figure~\ref{fig:corr_field} displays the Pearson $r$-values between each field parameter and the respective wear indicators. The analysis focuses on \textit{field parameters}, which are derived from the entire profile using mathematical definitions. Particularly strong correlations are observed for several \textit{hybrid parameters}, notably $R_\mathrm{dq}$ (root mean square gradient), $R_\mathrm{dl}$ (developed length), and $R_\mathrm{dr}$ (developed length ratio), all of which consistently exhibit high positive correlation coefficients ($r = 0.88-0.92$) across all three wear metrics.

In addition to the previously discussed parameters, slightly lower but still significant correlation values were observed for the following:
\begin{itemize}
    \item Reduced peak height ($R_\mathrm{pk}$) with correlation values up to $0.8$,
    \item Arithmetic mean of the absolute slope ($R_\mathrm{da}$) also reaching values up to $0.8$,
    \item Autocorrelation length ($R_\mathrm{al}$) with negative correlations down to $-0.76$.
\end{itemize}

In contrast to field parameters, \textit{feature parameters} are not derived from the entire profile but from individual features extracted by different segmentation methods, for which the default settings were applied here. Among the default feature parameters shown in Fig.~\ref{fig:corr_feature}, only the density of peaks ($R_\mathrm{pd}$) and the density of pits ($R_\mathrm{vd}$) stand out, both showing correlations up to $0.76$.

Except for $R_\mathrm{pk}$, these correlations indicate that the gradient- and length-related characteristics of the profile grow with wear. Whether this is due to changes in the deterministic groove shape or from an increasing stochastic microstructure cannot yet be distinguished at this level and is addressed in Section~\ref{sec:mean_feature}.

\section{Feature characterization}\label{sec:tool_grooves}
Feature characterization allows features of a topography to be extracted using watershed segmentation and statistically evaluated. For areal topographies, it was already standardized in ISO 25178-2 \cite{25178-2} in 2012 and was transferred to the ISO 21920-2 \cite{21920-2} standard for profile topographies in 2021. The underlying watershed segmentation is standardized in ISO 16610-85 \cite{16610-85} for areal and ISO 16610-45 \cite{16610-45} for profile topographies. An implementation guideline for feature characterization for profiles is described in detail in \cite{Müller2025}. For the following chapters, the five steps of feature characterization are illustrated in Fig.~\ref{fig:steps} and briefly summarized below:
\setlength{\leftmargini}{12pt}
\begin{enumerate}
 \item \textbf{Watershed segmentation}: The profile is partitioned into dales (between local maxima) or, by mirroring, into hills, following ISO 16610-45 \cite{16610-45}. In the remainder, ``D''/``V'' denote dale/pit and ``H''/``P'' denote hill/peak segmentation.
\item \textbf{Pruning}: To avoid oversegmentation, small or irrelevant segments are merged with neighboring larger ones. Pruning can be performed by one of four attributes: feature height (``Wolf pruning''), width (``Width''), volume (``Volume''), or developed length (``DevLength''). For Wolf pruning and width, the threshold may also be given as a percentage of $R_\mathrm{z}$ or of the evaluation length $l_\mathrm{e}$.
\item \textbf{Significant features}: Only features declared significant enter the statistical evaluation. ``All'' keeps every feature, while ``Open'' and ``Closed'' keep features whose water level lies below or above a nesting index $\mathit{N\!I}_{\mathrm{sig}}$, which may also be given as a percentage. ``Top'' and ``Bottom'' keep the $\mathit{N\!I}_{\mathrm{sig}}$ highest peaks or lowest pits.
\item \textbf{Feature attributes}: Geometric parameters such as height (``HDh''), width (``HDw''), volume (``HDv''), length (``HDl''), peak height/pit depth (``PVh''), or curvature (``Curvature'') are determined for each significant feature. These values describe the shape and size of the respective surface features. 
\item \textbf{Attribute statistics}: Finally, statistical parameters (e.g., mean value, standard deviation, density) are derived from the individual attribute values.
\end{enumerate}

It should be noted that ISO 21920-2 \cite{21920-2} explicitly conceives this method as open and extensible: the listed tools (feature attributes, significance filters, pruning criteria, and attribute statistics) are not exhaustive but may be supplemented by further ones tailored to the specific application. Accordingly, the present work also employs tools that deviate from those defined in the standard, in particular a custom significance filter and additional feature attributes.

The feature characterization is now adapted to the turning process by incorporating a priori knowledge. The repetitive grooves left by the cutting edge of the tool are recognizable to the human eye and potentially carry information about the tool wear. The watershed segmentation is therefore parameterized to extract these grooves specifically.

\begin{figure}
    \centering
    \includegraphics[scale=1]{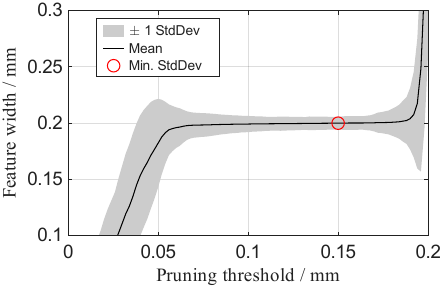}
    \caption{Influence of the width-based pruning threshold on the segmentation of the tool grooves. The mean feature width forms a plateau around \SI{0.2}{\milli\metre}, corresponding to the feed rate, and the standard deviation reaches its minimum at approximately \SI{0.15}{\milli\metre}.}     
    \label{fig:Threshold_analysis}
\end{figure}

\begin{figure}
    \centering
    \includegraphics[scale=1]{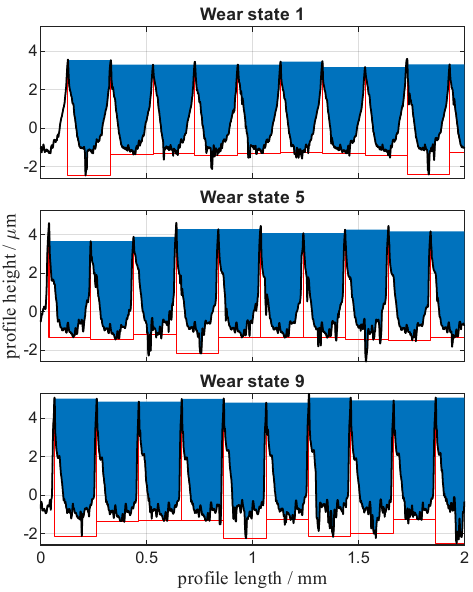}
    \caption{Watershed segmentation of the tool grooves.}     
    \label{fig:profiles_watershed}
\end{figure}

To determine a suitable pruning threshold, the feature width and its standard deviation were evaluated across all measured profiles for width-based pruning thresholds ranging from \SI{0}{\milli\metre} to the feed rate of \SI{0.2}{\milli\metre} (see Fig.~\ref{fig:Threshold_analysis}). A pronounced plateau of the mean feature width forms at the feed rate, indicating that the segmentation consistently captures the tool grooves over a broad range of thresholds. The minimum standard deviation, indicating the most stable segmentation, was found at approximately \SI{0.15}{\milli\metre} and was therefore selected as the width-based pruning threshold. It is combined with a significance filter that excludes features with a width below \SI{0.19}{\milli\metre} or above \SI{0.21}{\milli\metre} as segmentation artifacts, corresponding to approximately the \SI{95}{\percent} confidence interval of the feature width distribution at this threshold. Consequently, less than \SI{1.8}{\percent} of the features are discarded, which reflects both the robustness of the segmentation and the high regularity of the machined surfaces, whose groove widths deviate only marginally from the feed rate. Figure~\ref{fig:profiles_watershed} shows three profiles of different wear states with the described segmentation as an example for one tool.

\begin{figure}
    \centering
    \includegraphics{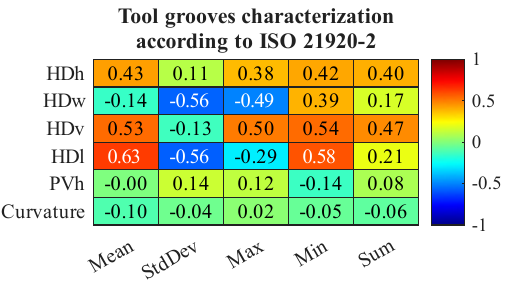}
    \caption{Pearson correlation of default feature parameters evaluated at groove level.}     
    \label{fig:correlations_grooves_feature_paramter}
\end{figure}
\begin{figure}
    \centering
    \includegraphics{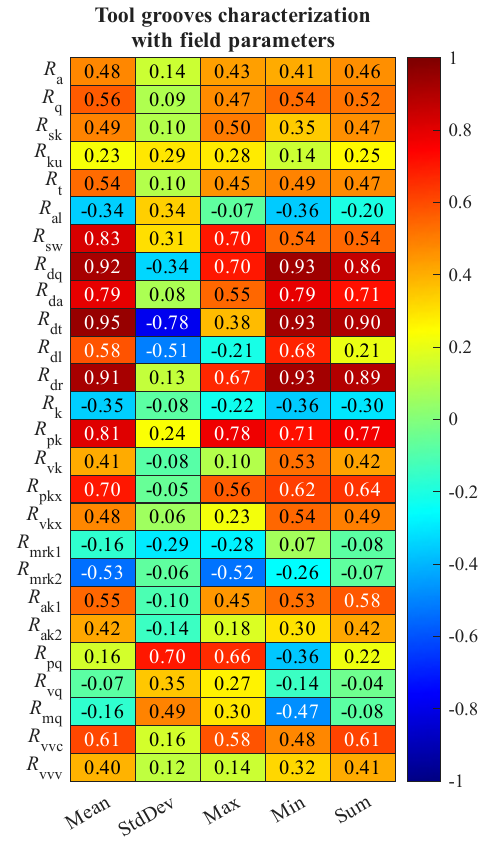}
    \caption{Pearson correlation of standardized field parameters evaluated at groove level.}
    \label{fig:correlations_tool_grooves}
\end{figure}

The groove-level evaluation is then carried out in two stages. First, the feature attributes defined in ISO 21920-2 were calculated, and the corresponding statistical values determined. These parameters were then correlated with the wear indicators. The resulting correlation values are shown in Fig.~\ref{fig:correlations_grooves_feature_paramter}. From this figure onward, the heatmaps display only the strongest correlation per parameter (i.e.\ the maximum $|r|$ across the three wear indicators $A_\mathrm{C}$, $\overline{V}_\mathrm{B}$, and $t_\mathrm{c}$) to keep them compact. However, compared to the strong correlations observed previously, the values are relatively weak. Second, the standardized field parameters are evaluated separately within each groove (Fig.~\ref{fig:correlations_tool_grooves}). The parameters reaching $|r| > 0.8$ are discussed below:
\begin{itemize}
    \item The mean dominant spatial wavelength $R_\mathrm{sw}$ now shows a strong correlation of 0.83. When viewed across the entire profile, the correlation remains close to 0, as the feed rate of \SI{0.2}{\milli\metre} was recorded across all wear states. Within the tool grooves, it is evident here that the short-wavelength structure exhibits a dominant wavelength that increases with wear.
    \item As with the field parameters for the entire profile, strong correlations are observed for the hybrid parameters, which, as expected, do not differ significantly. However, the correlation coefficient for the maximum absolute gradient $R_\mathrm{dt}$ has increased to the previous maximum of 0.95. This can be explained by the fact that the maximum absolute gradient in the overall profile likely occurs rather randomly somewhere in the profile due to outliers. Meanwhile, when averaged across the grooves, there is a strong correlation with wear. The dispersion of the parameter values also appears to decrease significantly with wear. The formal feature characterization descriptor according to ISO~21920-2 reads: FC; D; Width \SI{0.15}{\milli\metre}; HDw \SIrange{0.19}{0.21}{\milli\metre}; $R_\mathrm{dt}$; Mean. In the following, this parameter is denoted as $\overline{R_\mathrm{dt}}_\mathrm{groove}$.
\end{itemize}

\section{Mean feature approach} \label{sec:mean_feature}
While the groove-level analysis presented in the previous section already provides a more comprehensive insight into wear-induced surface changes, it is not yet clear whether the strong correlations are caused by changes in groove shape or by changes in the microstructure within the grooves. Therefore, an approach based on a mean feature is introduced, which allows the profile $z$ to be decomposed into a deterministic and a stochastic component, $z = z_{\det} + z_{\mathrm{stoch}}$.

\begin{figure}
    \centering
    \includegraphics{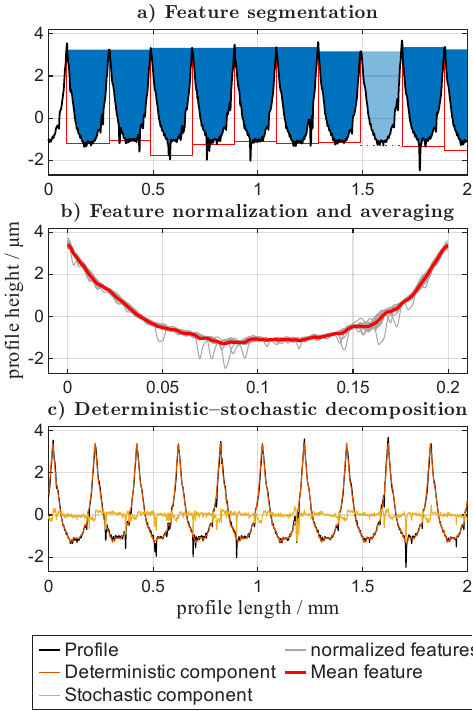}
    \caption{Decomposition of a roughness profile into deterministic and stochastic components using the mean feature approach: a) segmentation and filtering of tool grooves, b) normalization and averaging to obtain the mean feature, and c) reconstruction of the deterministic profile and extraction of the stochastic residual.}
    \label{fig:mean_feature_process}
\end{figure}
\begin{figure}
    \centering
    \includegraphics{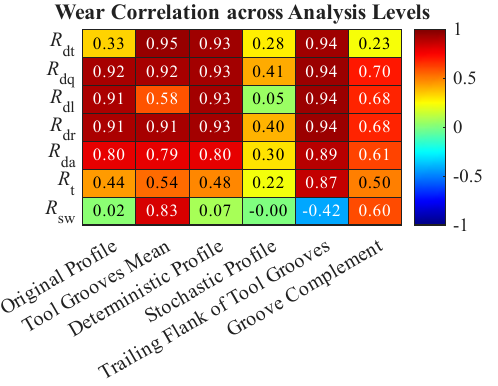}
    \caption{Comparison of Pearson correlation coefficients with wear across six analysis strategies: original profile, groove-level mean, deterministic profile, and stochastic profile. Only parameters with $|r| > 0.8$ in at least one category are shown.}
    \label{fig:correlations_comparing_approaches}
\end{figure}

Therefore, the tool grooves are segmented as described above (see Fig.~\ref{fig:mean_feature_process}(a)). In the next step, the significant features are normalized to a uniform width equal to the feed rate of \SI{0.2}{\milli\metre} and resampled to a common number of points by shape-preserving piecewise cubic interpolation (PCHIP). Since the significance filter retains only features whose width lies within \SI{0.19}{\milli\metre} and \SI{0.21}{\milli\metre}, this rescaling amounts to at most about \SI{5}{\percent} and therefore introduces no relevant shape distortion. Because of the uniform width, letting the features start at the same position is sufficient to align them. The vertical offset is removed by subtracting the mean of each feature. The normalized features are finally averaged point by point to yield the representative groove shape (see Fig.~\ref{fig:mean_feature_process}(b)).

A deterministic profile can now be generated from this mean feature by concatenating the mean features and realigning them with the original profile via cross-correlation. Since the grooves in the original profile have slightly different widths, the deterministic profile is adapted to the respective original width groove by groove through local stretching or compression. This ensures that the subsequently formed difference contains exclusively stochastic profile components and is not distorted by width differences (see Fig.~\ref{fig:mean_feature_process}(c)).

In contrast to alternative approaches that rely on a fixed period and process the entire profile, such as time-synchronous averaging~\cite{Braun2011}, Fourier representations~\cite{Groche2023}, or Gaussian-process models of the surface~\cite{Yan2024,Jawaid2025}, the proposed feature-based averaging operates on individually segmented features. This makes it possible to discard malformed or defective features before averaging so that they do not contaminate the resulting mean shape. Here, this defect criterion is realized by the significance filter of the watershed segmentation introduced in Section~\ref{sec:tool_grooves}, which excludes features whose width deviates from the feed rate. The approach is not restricted to strictly periodic surfaces: any class of features that can be segmented, for instance, individual abrasive marks or chatter structures, can be averaged in the same way regardless of their spatial regularity.

Figure~\ref{fig:correlations_comparing_approaches} compares the correlations between the parameters and wear for six analysis levels, the last two of which (columns 5 and 6) are introduced in detail only later in this section. Only parameters with a correlation coefficient greater than 0.8 in at least one category are shown. It can be seen that the strong correlations are almost entirely driven by the deterministic profile component, whose correlation coefficients range from 0.78 to 0.93 and are comparable to those of the groove-based analysis. The stochastic component, on the other hand, consistently exhibits weak correlations ($r \leq 0.41$). This confirms that the wear-induced changes in surface topography are primarily determined by changes in groove shape and thus in cutting edge geometry, rather than by changes in the stochastic microstructure within the grooves.

\begin{figure*}
\begin{tabular}{p{0.48\textwidth} p{0.48\textwidth}}
  \vspace{0pt} \includegraphics[width=0.98\linewidth]{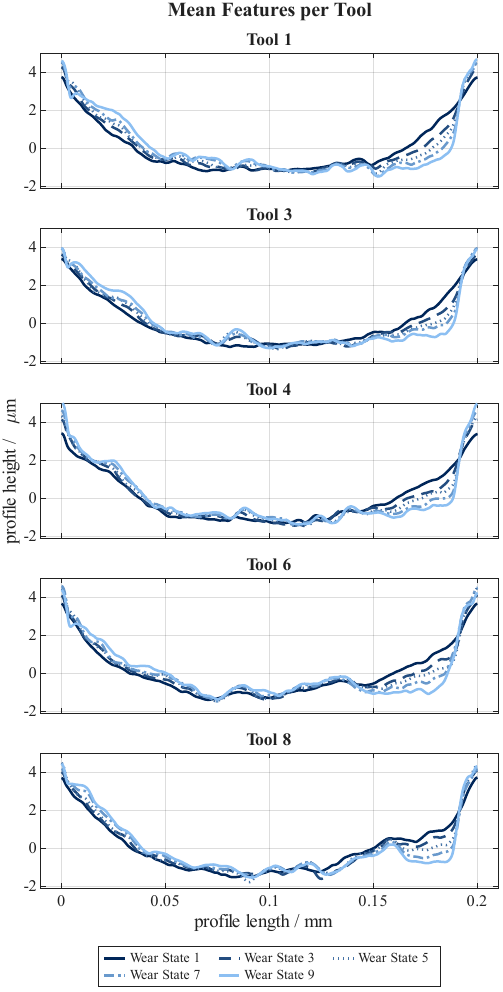} &
  \vspace{0pt} \includegraphics[width=0.98\linewidth]{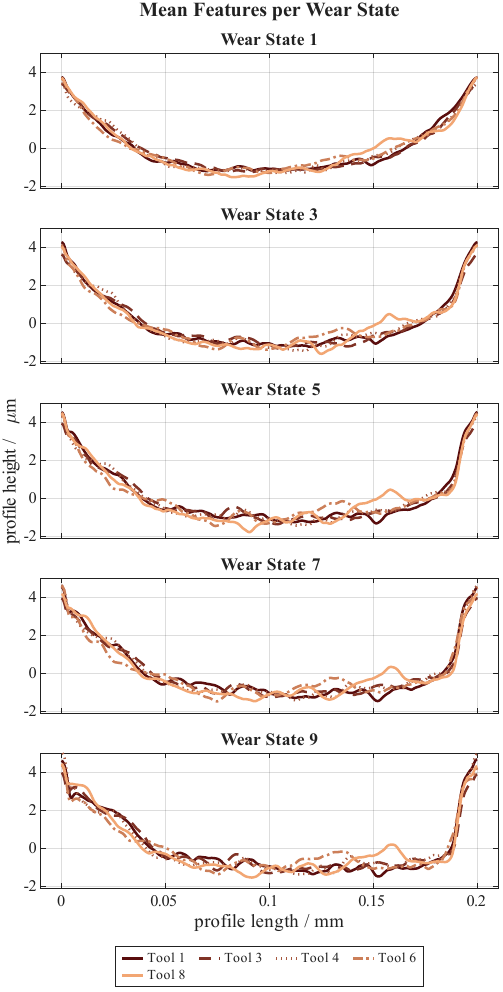}
\end{tabular}
\caption{Mean features per tool and per wear state.}
\label{fig:mean_features}
\end{figure*}

Since the mean feature represents this groove shape, it can be used to visualize wear-induced changes. Figure~\ref{fig:mean_features} is organized in two columns: the left column shows the progression of groove shapes for a single tool across multiple wear states, and the right column compares five different tools at corresponding wear levels, indicating whether wear-induced changes are consistent across tools or exhibit tool-specific variations. The remaining six tools are shown in Fig.~\ref{fig:mean_period_appendix} in Appendix~\ref{appendix:figures}.
As wear progresses, various trends can be observed across all tools:
\begin{itemize}
    \item An increasing steepening of the flank forms in the middle to right region of the feature.
    \item This steepening is accompanied by a deepening depression in the adjacent valley.
    \item The left side becomes slightly flatter.
    \item While the initial wear state still exhibits a comparatively symmetrical groove shape, this deviates successively as wear progresses.
\end{itemize}

Although the changes are very similar, they differ slightly in their characteristics, as can be clearly seen in the right-hand column of Fig.~\ref{fig:mean_features}. Only the trailing flank, from approximately \SIrange{0.17}{0.2}{\milli\metre}, changes very uniformly and remains virtually identical, which may be the reason for the strong correlations with gradient-based parameters.

To prove this, the grooves were segmented analogously to Section~\ref{sec:tool_grooves}, and then the field parameters were calculated separately for the trailing flank (the last \SI{0.03}{\milli\metre} of the feature, corresponding to the \SIrange{0.17}{0.2}{\milli\metre} range identified above as changing uniformly) and for the remaining part of the groove and averaged across all grooves. The resulting correlations are shown in Fig.~\ref{fig:correlations_comparing_approaches} in columns 5 and 6. It can be seen that the correlation values of the trailing flank almost match those at the groove level. This confirms the hypothesis: the trailing flank is the dominant carrier of wear-induced profile changes. The remaining part of the groove, on the other hand, exhibits significantly weaker correlations ($r \leq 0.70$), suggesting that this area undergoes less systematic changes as wear progresses.
The total height parameter $R_\mathrm{t}$ now also shows a strong correlation, reflecting the aforementioned deepening.

It has been shown that the strong correlations are dominated by changes in the deterministic component of the surface, and particularly by the trailing flank of the profile. This suggests that these changes directly reflect the cutting edge geometry, which can therefore be reconstructed from the workpiece profile. This will be validated in Section~\ref{sec:linking}.

\section{Linking surface profile to cutting edge geometry} \label{sec:linking}
\begin{figure}
    \centering
    \includegraphics{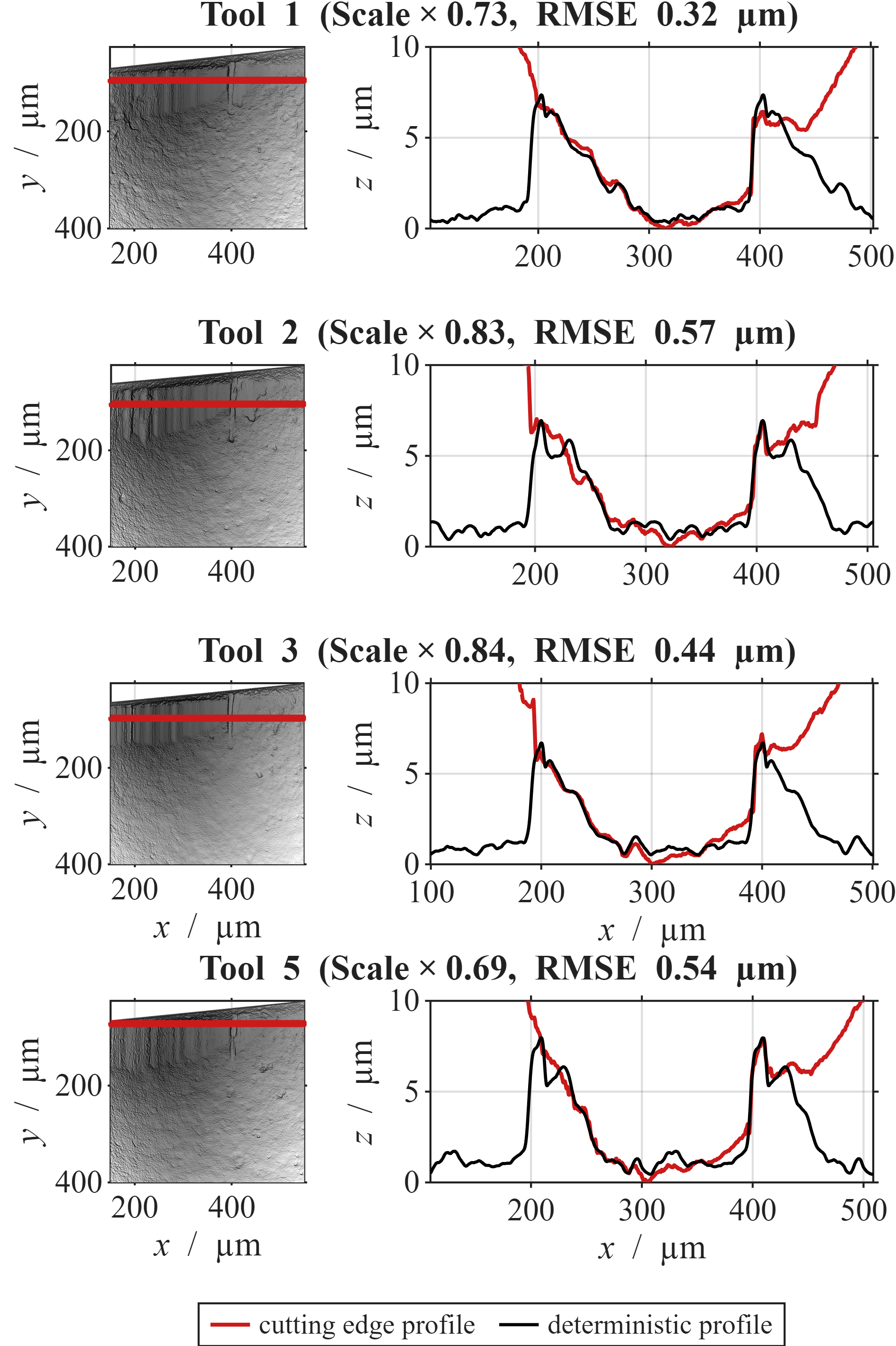}
    \caption{Comparison of the deterministic profile reconstructed from the workpiece roughness measurement and the band-averaged cutting edge profile obtained from the confocal measurement of the tool for four tools. Left: confocal topography of the cutting edge with the extracted band position (red). Right: overlay of the deterministic profile (black) and the cutting edge profile (red). Each subplot title gives the multiplicative scale factor applied to the cutting edge profile as the ×-value, followed by the root-mean-square deviation between the two profiles.}
    \label{fig:Compare_profiles}
\end{figure}
The aim of this section is to provide direct evidence that the mean feature not only correlates with individual wear indicators but also reconstructs the engaged section of the cutting edge geometry from the workpiece profile. To this end, the deterministic profile obtained from the surface is compared with a confocal measurement of the cutting edge of the same tool.

Since the exact location of the effective cutting edge profile cannot be unambiguously determined due to measurement inaccuracies, alignment errors, and deviations during tool clamping, the deterministic profile reconstructed from the mean feature was systematically shifted along the cutting edge to identify the position with the best match. At this point, the cutting edge profile from the confocal measurement was averaged over a \SI{5}{\micro\metre} wide band in the cutting direction. Additionally, a multiplicative scaling factor and a vertical offset were allowed. The lateral position, the scaling factor, and the offset were determined per tool by a least-squares fit minimizing the mean squared deviation between the two profiles. The resulting scaling factor is stated above each sub-image in Fig.~\ref{fig:Compare_profiles}. Scaling factors lower than 1 are due to the elastic spring-back of the workpiece.
The remaining tools are shown in Fig.~\ref{fig:Compare_profiles_appendix} in Appendix~\ref{appendix:figures}.\footnote{Individual tools were no longer available for this study, as they had been destroyed during microstructural analyses.}

As shown in Fig.~\ref{fig:Compare_profiles}, the two profiles are in close agreement, with a root-mean-square deviation of \SI{0.50}{\micro\metre} on average across all tools (range \SIrange{0.32}{0.85}{\micro\metre}). The upper end of this range corresponds to Tool~10 (Fig.~\ref{fig:Compare_profiles_appendix}), whose secondary cutting edge exhibits a localized spall on the flank face. This singular damage feature is absent from the recurrent groove shape captured by the mean feature and therefore increases the local mismatch between the two profiles, which accounts for its comparatively high deviation rather than indicating a shortcoming of the method.
Thus, the direct comparison demonstrates that the wear-related changes observed in the workpiece profile directly reflect the geometry of the tool cutting edge, and that the mean feature therefore allows conclusions to be drawn about the engaged section of the cutting edge. The steepening of the trailing flank observed across all tools can now be directly attributed to notch wear on the secondary cutting edge. 
If sufficiently pronounced, the notch may also protrude into the previously machined turning groove, which can be seen, for example, in Fig.~\ref{fig:Compare_profiles} for tools 2 and 5.

This wear mechanism is well documented for the turning of hardened steels. It is described as a localized indentation at the worn secondary cutting edge, arising from combined mechanical, thermal, and abrasive edge loading, that is directly transferred to the machined surface~\cite{Klocke2018, Jochmann2001}. For hard turning with mixed-ceramic tools, Grzesik~\cite{Grzesik2008} showed that the notch forming at the secondary cutting edge is copied onto the workpiece profile, transforming the originally blunt profile tips into sharp tips on the trailing flank, an effect he termed the \textit{profile sharpening effect}. Exactly this effect is observed in the profiles examined here. Consistent with this, Derani et al.~\cite{Derani2021} report that a gradient-based descriptor of the turned profile follows flank wear far more closely than the amplitude parameter $R_\mathrm{a}$.
The contribution of the present work lies in the method rather than in the phenomenon itself. Whereas the cited studies capture the transfer of notch wear onto the profile only qualitatively or through ad hoc, self-defined descriptors, the present work employs the standardized feature characterization of ISO~21920-2 to visualize the wear-induced changes via the mean feature, to demonstrate the imaging of the cutting edge, and to quantify the changes through a single standardized parameter.

These findings cannot be generalized without restriction. The formation and geometry of notch wear depend on workpiece material and hardness, tool substrate and coating, cutting edge geometry, and process parameters~\cite{Klocke2018}. In other material systems, like titanium and nickel alloys, the dominant wear mechanisms and their imprint on the surface profile can differ substantially~\cite{Liang2019}. The strong correlation between $\overline{R_\mathrm{dt}}_\mathrm{groove}$ and tool wear established in this study is therefore specific to the investigated combination (see Section~\ref{sec:method}), and transferability to other configurations requires further investigation.

\section{Evaluation of additional characterization and modeling approaches}\label{sec:additional}
The preceding sections have demonstrated that the groove-level maximum absolute gradient $\overline{R_\mathrm{dt}}_\mathrm{groove}$ is the single most informative parameter for all three wear indicators simultaneously, with Pearson correlation coefficients of $0.91$, $0.92$, and $0.95$ for $A_\mathrm{C}$, $\overline{V}_\mathrm{B}$, and $t_\mathrm{c}$, respectively. To assess whether more elaborate strategies can improve upon this finding, two questions are addressed. First, whether any other single parameter outperforms it, examined both by an exhaustive, data-driven search through the feature characterization toolbox and by a set of custom groove-shape descriptors. Second, whether combining several parameters in a multivariate model improves the prediction. As shown below, neither direction yields a substantial improvement over the simple linear relationship with $\overline{R_\mathrm{dt}}_\mathrm{groove}$.

\subsection{Exhaustive search for a superior single parameter}\label{sec:single_param_search}
The first search is exhaustive and incorporates no prior process knowledge, using only the standardized criteria of the ISO~21920-2 toolbox. The pruning threshold for Wolf pruning was varied as a percentage of $R_\mathrm{z}$ in \SI{2}{\percent} increments from \SI{2}{\percent} to \SI{98}{\percent}, the width-based pruning threshold as a percentage of the evaluation length $l_\mathrm{e}$ in \SI{0.1}{\percent} increments from \SI{0.1}{\percent} to \SI{10}{\percent}, and the significance criterion (``Open''/``Closed'') as a percentage of the inverse material ratio in \SI{2}{\percent} increments. Combined with all remaining toolbox options (feature type, pruning type, feature attributes, and attribute statistics), this yields \num{920700} parameters per profile. 
Of these, only 190 reach $|r| \geq 0.8$. The strongest toolbox parameter reaches a correlation of at most 0.85, markedly below $\overline{R_\mathrm{dt}}_\mathrm{groove}$.

The second search adds custom groove-shape descriptors beyond the standardized attributes of Section~\ref{sec:tool_grooves}. Each descriptor was evaluated statistically (mean, maximum, minimum, standard deviation, and sum) across all grooves per profile. The strongest absolute correlation it attains with any of the three wear indicators ($A_\mathrm{C}$, $\overline{V}_\mathrm{B}$, $t_\mathrm{c}$) is reported below:
\begin{itemize}
    \item secant-based gradient on the trailing flank ($|r| = 0.93$),
    \item groove symmetry ($|r| = 0.92$),
    \item lateral centroid position of the groove cross-section ($|r| = 0.84$),
    \item skewness of the groove-shape distribution ($|r| = 0.79$),
    \item fitted circle radius as a measure of groove curvature ($|r| = 0.73$).
\end{itemize}
Several of these descriptors correlate strongly with all three wear indicators, but they also correlate strongly with $\overline{R_\mathrm{dt}}_\mathrm{groove}$ itself, and none reaches a higher correlation with the wear indicators than $\overline{R_\mathrm{dt}}_\mathrm{groove}$.

Neither search yields a single parameter that correlates more strongly with the wear indicators than the physically motivated choice.

\subsection{Multivariate models}\label{sec:multivariate}
The validation strategy is critical for the reliability of the reported metrics. Since the dataset consists of twelve tools with nine wear states and three replicate measurements each, a random split of the samples would cause replicate measurements of the same tool to appear simultaneously in the training and test dataset. The resulting information leakage leads to optimistically biased performance metrics. All models are therefore evaluated using a Leave-One-Tool-Out cross-validation (LOTO): in each of the twelve iterations, exactly one tool with all its associated measurements serves as the test set, while the remaining eleven tools train the model. The reported coefficient of determination $R^2_\mathrm{LOTO}$ is computed from the pooled predictions of all held-out tools and thus estimates generalization to a previously unseen tool.

The argument proceeds in three steps: an unguided data-driven selection (does the search rediscover the physical parameter?), a residualized selection (does anything remain beyond it?), and a predictive model comparison (can a multivariate model exploit more, including nonlinearly?).

The complete parameter pool of \num{921655} features assembled previously, comprising the \num{920700} toolbox parameters of Section~\ref{sec:single_param_search}, the \num{810} custom groove-shape descriptors, and \num{145} standardized parameters evaluated on the original as well as the deterministic and stochastic profiles, was subjected to a leakage-free, nested stability selection \cite{Meinshausen2010,Shah2013}, detailed in Appendix~\ref{appendix:stable}. In essence, an Elastic Net is repeatedly fitted to tool-wise subsamples within each training partition, and only parameters selected with high frequency across the subsamples are retained.

This unguided search converges on the physically motivated choice. The seven parameters retained for all three wear indicators simultaneously are, without exception, gradient, symmetry, or height-span descriptors evaluated on the very same groove segmentation as $\overline{R_\mathrm{dt}}_\mathrm{groove}$ (Appendix~\ref{appendix:stable}). The data-driven selection and the physical derivation of Section~\ref{sec:linking} thus arrive at the same low-dimensional structure. Consistently, this set is strongly collinear, with variance inflation factors (VIF) reaching 17 ($A_\mathrm{C}$), 19 ($\overline{V}_\mathrm{B}$), and 37 ($t_\mathrm{c}$): the selected parameters do not represent several independent influences but fundamentally the same wear-induced steepening of the trailing groove flank already captured by $\overline{R_\mathrm{dt}}_\mathrm{groove}$.

To test whether any parameter carries information \emph{beyond} $\overline{R_\mathrm{dt}}_\mathrm{groove}$, the identical stability selection was repeated with $\overline{R_\mathrm{dt}}_\mathrm{groove}$ entered as a fixed, unpenalized covariate. This is equivalent to selecting on the residuals of the single-parameter fit, with the single-parameter fit re-estimated within each subsample and each training partition so that the residualization never uses the held-out tool. For $\overline{V}_\mathrm{B}$ and $t_\mathrm{c}$, no parameter is selected stably, and for $A_\mathrm{C}$ three parameters cross the threshold but prove to be a non-transferable in-sample artefact (Appendix~\ref{appendix:stable}). No parameter provides robust incremental information beyond $\overline{R_\mathrm{dt}}_\mathrm{groove}$.

The incrementally unstable parameters might still be exploited jointly by a model. To test this, three models with fundamentally different approaches to collinearity and nonlinearity are compared for each wear indicator ($A_\mathrm{C}$, $\overline{V}_\mathrm{B}$, and $t_\mathrm{c}$). Multiple linear regression, estimated by ordinary least squares (OLS), does not address collinearity, Partial Least Squares (PLS)~\cite{Wold2001} projects the parameters onto a few uncorrelated latent components, and Gradient-Boosted Trees (XGBoost)~\cite{Chen2016} can model nonlinear relationships and interactions. A simple linear regression with $\overline{R_\mathrm{dt}}_\mathrm{groove}$ as the sole predictor serves as the reference. Table~\ref{tab:model_comparison} summarizes the results.

\begin{table}[h]
\centering
\caption{Comparison of cross-validated coefficients of determination $R^2_\mathrm{LOTO}$ (Leave-One-Tool-Out) for the investigated modeling approaches and wear indicators. The reference uses exclusively the groove-level mean maximum absolute gradient $\overline{R_\mathrm{dt}}_\mathrm{groove}$ as the single predictor.}
\label{tab:model_comparison}
\begin{tabular}{l c c c}
\hline
Model & $A_\mathrm{C}$ & $\overline{V}_\mathrm{B}$ & $t_\mathrm{c}$ \\
\hline
Reference ($\overline{R_\mathrm{dt}}_\mathrm{groove}$)  & 0.82 & 0.83 & 0.89 \\
Multiple linear regression (OLS)              & 0.81 & 0.85 & 0.91 \\
PLS (latent components)                       & 0.81 & 0.86 & 0.93 \\
XGBoost (nonlinear)                           & 0.82 & 0.84 & 0.91 \\
\hline
\end{tabular}
\end{table}

All three models yield nearly identical coefficients of determination, spanning a range of only approximately $0.03$ and barely exceeding the single-parameter reference. This result is informative from three directions. First, OLS performs only marginally worse than PLS despite the high collinearity. The latent projection stabilizes the linear model but does not extract any additional signal. Second, the nonlinear XGBoost does not outperform the linear approaches, so no relevant nonlinear relationships are present. Third, a PLS variant with explicitly added pairwise interaction terms changes the values only within $\pm 0.01$ ($0.83 / 0.87 / 0.92$), so no exploitable interaction signal can be demonstrated either.

A paired, tool-wise cluster bootstrap~\cite{Field2007} confirms that for none of the three indicators does the 95\,\% confidence interval of the improvement over the reference exclude zero, so the multivariate models provide no significant added value. The close agreement between the in-sample and LOTO coefficients of the reference (gap at most $0.013$) further indicates that the unexplained variance is essentially measurement noise.

These results confirm that $\overline{R_\mathrm{dt}}_\mathrm{groove}$ at the groove level captures the essential wear-induced signal. Multivariate models yield only a marginal, statistically non-significant improvement that does not justify the considerably higher modeling effort.

\section{Conclusion and future work}\label{sec:conclusion}
This study investigated the extent to which the feature characterization standardized in ISO~21920-2 can contribute to the description of wear-induced changes in turned surfaces. A correlation analysis of the standardized field and feature parameters shows that the hybrid parameters $R_\mathrm{dq}$, $R_\mathrm{dl}$, and $R_\mathrm{dr}$ correlate strongly ($r = 0.88$--$0.92$) with the wear indicators $A_\mathrm{C}$, $\overline{V}_\mathrm{B}$, and $t_\mathrm{c}$. When watershed segmentation is additionally used to isolate the tool grooves, the groove-level mean maximum absolute gradient $\overline{R_\mathrm{dt}}_\mathrm{groove}$ emerges as the single most informative parameter, reaching $r = 0.95$.

The mean-feature approach decomposes the roughness profile into a deterministic and a stochastic component using a mean groove feature. The wear-induced changes are almost entirely carried by the deterministic component, while the stochastic component correlates only weakly with wear ($r \leq 0.41$). A separate evaluation of the trailing flank of the groove and the remaining groove section further localizes the wear-induced changes to the trailing flank. A direct comparison of the mean feature with confocal measurements of the cutting edge confirms that the mean feature reconstructs the engaged section of the cutting edge geometry and attributes the observed steepening of the trailing flank to notch wear on the secondary cutting edge. Since the dominant wear mechanisms vary substantially with the machining configuration~\cite{Klocke2018,Liang2019}, the transferability of these findings, in particular the dominant role of the trailing flank, requires further investigation.

An extensive evaluation of over \num{920000} feature characterization combinations, custom groove-shape descriptors, and multivariate regression models showed that none of these approaches substantially improve upon a simple linear regression with the groove-level mean maximum absolute gradient $\overline{R_\mathrm{dt}}_\mathrm{groove}$, which alone explains 83--90\,\% of the wear variance ($R^2_\mathrm{LOTO} = 0.82$--$0.89$ under leave-one-tool-out cross-validation). A single, physically motivated parameter is therefore sufficient for wear estimation that remains robust across the tool population. Beyond wear monitoring, the functional performance of a machined surface (for instance in sealing, friction, or load-bearing contact) is governed by individual structural features rather than by field parameters averaged over the entire profile, so that the feature-resolved analysis pursued here offers a natural and promising framework for functionally motivated surface characterization in general.

This gives rise to two immediate points of interest for future work. First, the fact that gradient-based parameters show high correlation coefficients with the wear indicators for this machining setup opens up the possibility of monitoring the wear inline using scattered light sensors. Second, the obtained, spatially resolved information on wear-induced profile changes provides a suitable database for validating and further developing the finite element models of the machining process developed within the framework of the grey-box approach in other works by the authors ~\cite{Berndt2025}.

\printcredits

\section*{Declaration of competing interest}
The authors declare that they have no known competing financial interests or personal relationships that could have appeared to influence the work reported in this paper.

\section*{Generative AI}
During the preparation of this work the authors used large language models to improve readability and language. After using this tool, the authors reviewed and edited the content as needed and take full responsibility for the content of the published article.

\section*{Acknowledgements}
This work was funded by the Deutsche Forschungsgemeinschaft (DFG, German Research Foundation) -- project numbers 461839204 and 521380776 (research priority program SPP 2402).

\section*{Data availability}
Data will be made available on request.

\bibliographystyle{model1-num-names}

\bibliography{cas-refs}



\appendix
\section{Supplementary figures}\label{appendix:figures}

\begin{table*}[pos=t]
\centering
\small
\caption{Profile roughness parameters of ISO~21920-2 \cite{21920-2}. Field parameters are derived from the entire profile, feature parameters from individual segmented features.}
\label{tab:iso_parameters}
\begin{tabular}[t]{l l}
\hline
\multicolumn{2}{c}{\textbf{\normalsize Field parameters}} \\
\hline
\multicolumn{2}{l}{\textbf{Height parameters}} \\
$R_\mathrm{a}$        & arithmetic mean height \\
$R_\mathrm{q}$        & root mean square height \\
$R_\mathrm{sk}$       & skewness \\
$R_\mathrm{ku}$       & kurtosis \\
$R_\mathrm{t}$        & total height \\
$R_\mathrm{zx}(l)$    & maximum height per section \\
\multicolumn{2}{l}{\textbf{Spatial parameters}} \\
$R_\mathrm{al}(l)$    & autocorrelation length \\
$R_\mathrm{sw}$       & dominant spatial wavelength \\
\multicolumn{2}{l}{\textbf{Hybrid parameters}} \\
$R_\mathrm{dq}$       & root mean square gradient \\
$R_\mathrm{da}$       & arithmetic mean of absolute gradient \\
$R_\mathrm{dt}$       & maximum absolute gradient \\
$R_\mathrm{dl}$       & developed length \\
$R_\mathrm{dr}$       & developed length ratio \\
\multicolumn{2}{l}{\textbf{Material ratio functions}} \\
$R_\mathrm{ml}(c)$    & material length \\
$R_\mathrm{mc}(c)$    & material ratio \\
$R_\mathrm{cm}(p)$    & inverse material ratio \\
$R_\mathrm{hd}(c)$    & height density \\
$R_\mathrm{vm}(p)$    & material volume \\
$R_\mathrm{vv}(p)$    & void volume \\
\multicolumn{2}{l}{\textbf{Material ratio parameters}} \\
$R_\mathrm{mr}(p,d_\mathrm{c})$ & relative material ratio \\
$R_\mathrm{dc}(p,q)$  & material ratio height difference \\
\multicolumn{2}{l}{\textbf{Stratified surfaces, material ratio curve}} \\
$R_\mathrm{k}$        & core height \\
$R_\mathrm{pk}$       & reduced peak height \\
$R_\mathrm{vk}$       & reduced pit depth \\
$R_\mathrm{pkx}$      & maximum peak height \\
$R_\mathrm{vkx}$      & maximum pit depth \\
$R_\mathrm{mrk1}$     & material ratio of hills \\
$R_\mathrm{mrk2}$     & material ratio of dales \\
$R_\mathrm{ak1}$      & area of hills \\
$R_\mathrm{ak2}$      & area of dales \\
\multicolumn{2}{l}{\textbf{Stratified surfaces, material probability curve}} \\
$R_\mathrm{pq}$       & plateau root mean square deviation \\
$R_\mathrm{vq}$       & dale root mean square deviation \\
$R_\mathrm{mq}$       & material ratio at plateau-to-dale transition \\
\multicolumn{2}{l}{\textbf{Volume parameters}} \\
$R_\mathrm{vmp}(p)$   & hill material volume \\
$R_\mathrm{vmc}(p,q)$ & core material volume \\
$R_\mathrm{vvc}(p,q)$ & core void volume \\
$R_\mathrm{vvv}(p)$   & dale void volume \\
\hline
\end{tabular}
\hfill
\begin{tabular}[t]{l l}
\hline
\multicolumn{2}{c}{\textbf{\normalsize Feature parameters}} \\
\hline
\multicolumn{2}{l}{\textbf{Parameters based on peak heights and pit depths}} \\
$R_\mathrm{pt}$       & maximum peak height \\
$R_\mathrm{p}$        & mean peak height \\
$R_\mathrm{vt}$       & maximum pit depth \\
$R_\mathrm{v}$        & mean pit depth \\
$R_\mathrm{z}$        & maximum height \\
\multicolumn{2}{l}{\textbf{Parameters based on profile elements}} \\
$R_\mathrm{sm}$       & mean profile element spacing \\
$R_\mathrm{smx}$      & maximum profile element spacing \\
$R_\mathrm{smq}$      & standard deviation of profile element spacings \\
$R_\mathrm{c}$        & mean profile element height \\
$R_\mathrm{cx}$       & maximum profile element height \\
$R_\mathrm{cq}$       & standard deviation of profile element heights \\
$R_\mathrm{pc}$       & peak count parameter \\
\multicolumn{2}{l}{\textbf{Parameters based on feature characterization}} \\
$R_\mathrm{pd}$       & density of peaks \\
$R_\mathrm{vd}$       & density of pits \\
$R_\mathrm{mpc}$      & arithmetic mean peak curvature \\
$R_\mathrm{mvc}$      & arithmetic mean pit curvature \\
$R_\mathrm{5p}$       & five-point peak height\\
$R_\mathrm{5v}$       & five-point pit depth\\
$R_\mathrm{10z}$      & 10-point height\\
\hline
\end{tabular}
\end{table*}

\begin{figure*}
\begin{tabular}{p{0.5\textwidth} p{0.5\textwidth}}
  \vspace{0pt} \includegraphics[scale=0.95]{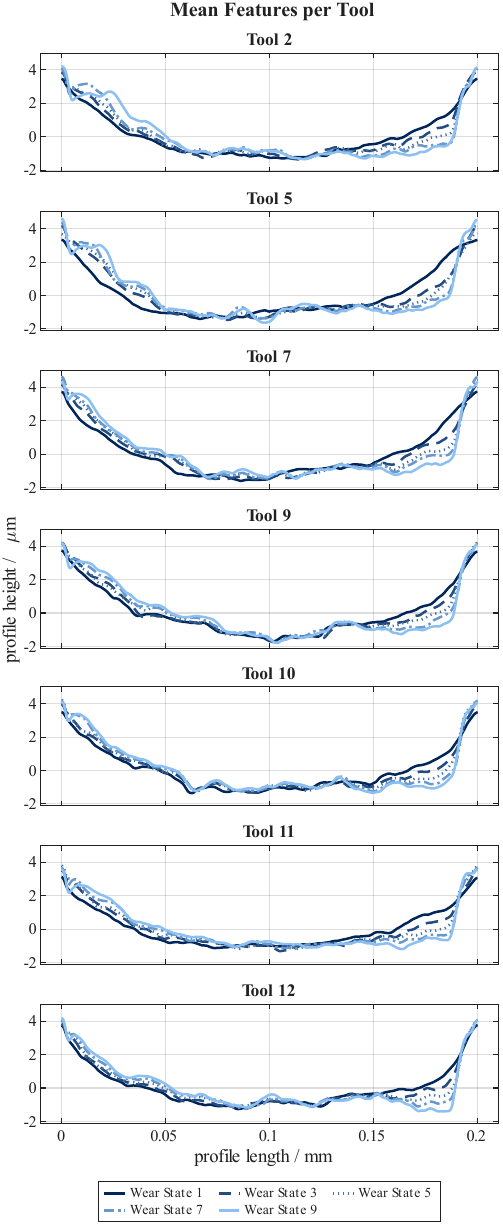} &
  \vspace{0pt} \includegraphics[scale=0.95]{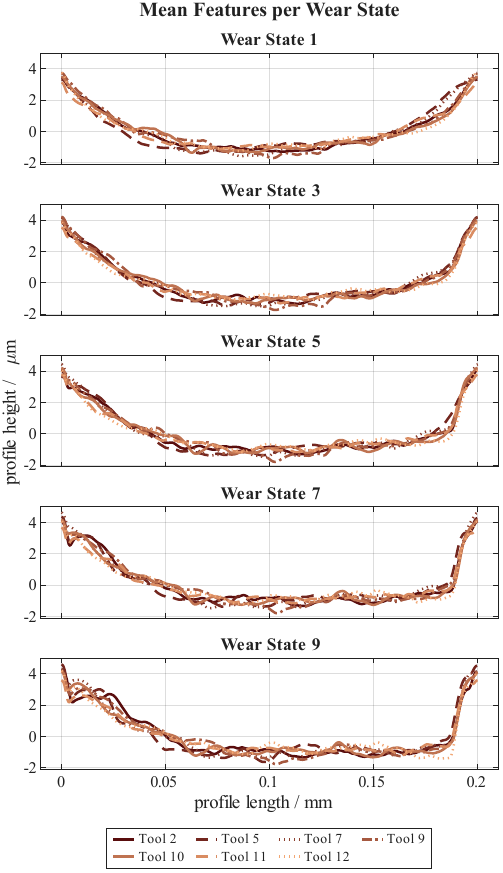}
\end{tabular}
\caption{Mean features per tool and per wear state.}
\label{fig:mean_period_appendix}
\end{figure*}

\begin{figure*}
    \centering
    \includegraphics{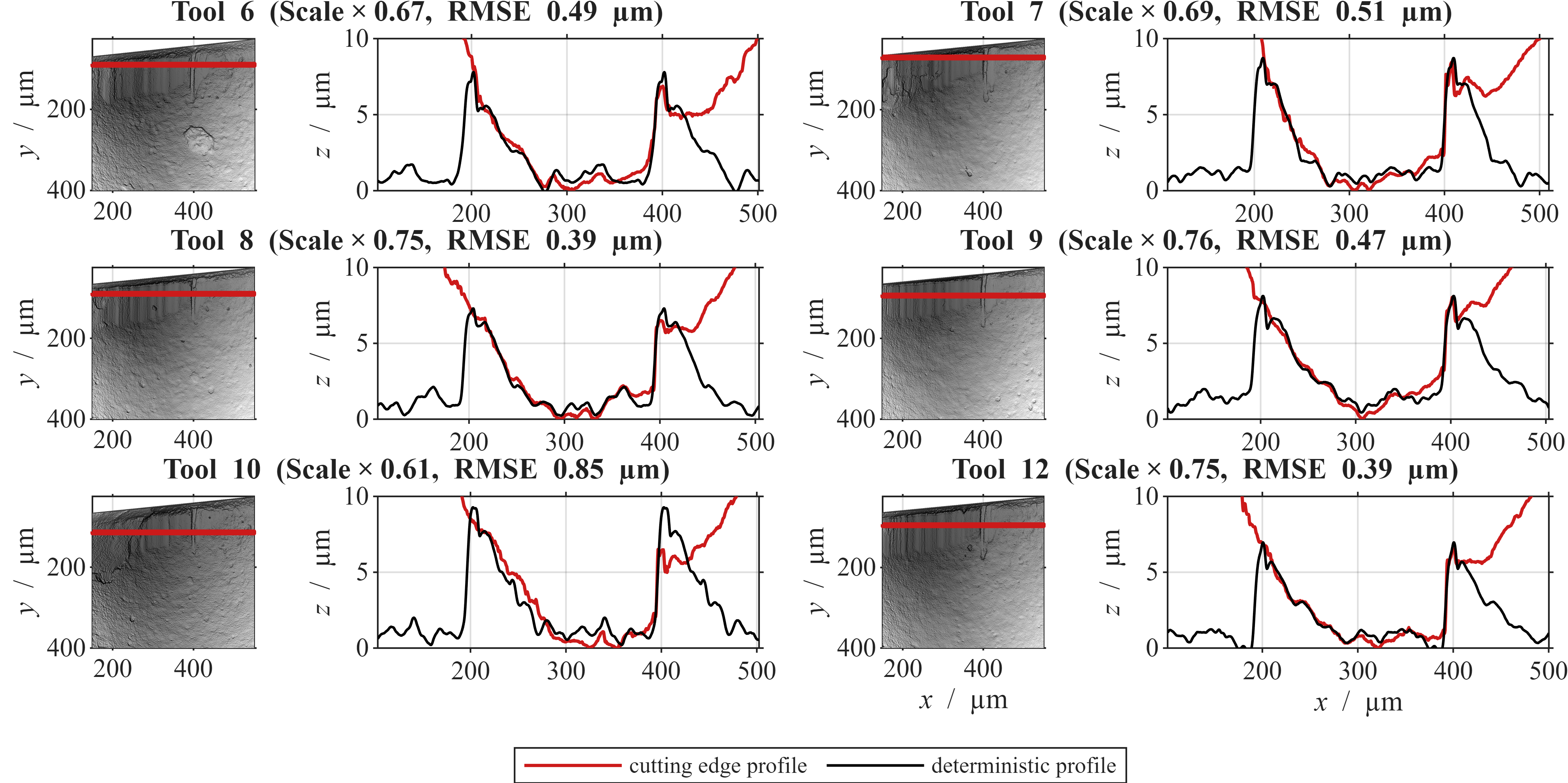}
    \caption{Comparison of the deterministic profile reconstructed from the workpiece roughness measurement and the band-averaged cutting edge profile obtained from the confocal measurement of the tool for the remaining tools. Left: confocal topography of the cutting edge with the extracted band position (red). Right: overlay of the deterministic profile (black) and the cutting edge profile (red). Each subplot title gives the multiplicative scale factor applied to the cutting edge profile as the ×-value, followed by the root-mean-square deviation between the two profiles.}
    \label{fig:Compare_profiles_appendix}
\end{figure*}
\clearpage
\section{Stability selection}\label{appendix:stable}

\subsection{Setup}
From the complete parameter pool, a purely unsupervised pre-filtering is first applied, i.e.\ without using the wear indicators, removing parameters with missing values, near-constant behavior, and near-identical, mutually highly correlated duplicates ($|r| \geq 0.95$). Since no target information is used in this step, it is leakage-free, and approximately \num{39000} features remain. The target-oriented stability selection \cite{Meinshausen2010,Shah2013} is then repeated within each training partition, so that the held-out tool enters neither the selection nor the training (nested validation). An Elastic Net estimator~\cite{Zou2005} ($\alpha = 0.5$, which groups correlated parameters) is fitted repeatedly to \num{100} subsamples drawn as \num{50} complementary pairs. The subsampling is performed tool-wise, splitting the twelve tools into two disjoint halves of six so that the group structure of the replicate measurements is preserved, consistent with the LOTO scheme. Rather than fixing a single regularization strength, the penalty is calibrated per subsample so that on the order of $q \approx 30$ parameters enter each fit. A parameter is retained if its selection frequency across the subsamples reaches the threshold $\pi_\mathrm{thr} = 0.6$. Under these settings the Meinshausen--Bühlmann bound $E[V] \leq q^2/((2\pi_\mathrm{thr}-1)\,p)$ limits the expected number of falsely selected parameters to well below one.

\subsection{Residualized selection}
For the residualized selection with $\overline{R_\mathrm{dt}}_\mathrm{groove}$ as a fixed covariate (Section~\ref{sec:multivariate}), the penalty calibration is unchanged, so on the order of $q \approx 30$ parameters still enter each individual fit, and the question is solely whether any of them is selected \emph{stably}. For $\overline{V}_\mathrm{B}$ and $t_\mathrm{c}$, no parameter reaches the threshold at all (maximum selection frequency $0.51$ and $0.52$, i.e.\ the floor expected for the complementary-pairs scheme in the absence of signal). For $A_\mathrm{C}$, three parameters formally cross the threshold, but they are three adjacent pruning-threshold variants of a single peak-to-valley-height parameter evaluated on the full profile, unrelated to the groove segmentation that carries the wear signal. They are reselected in at most one of the twelve leave-one-tool-out folds, and adding them to the reference lowers the cross-validated $R^2_\mathrm{LOTO}$ by $0.017$ rather than raising it. They are therefore an in-sample artefact, not transferable information. This also illustrates why the nested, tool-wise evaluation is decisive. The global Meinshausen--Bühlmann bound (here $E[V] \leq 0.1$) controls false positives within the analyzed sample, but it does not guarantee generalization to an unseen tool.

\subsection{Stable parameter set}
Table~\ref{tab:stable_set} lists the seven parameters that the stability selection of Section~\ref{sec:additional} retained for all three wear indicators simultaneously. The per-indicator sets comprise 10 ($A_\mathrm{C}$), 12 ($\overline{V}_\mathrm{B}$), and 11 ($t_\mathrm{c}$) parameters. All of them are evaluated on the groove segmentation introduced in Section~\ref{sec:tool_grooves}, i.e.\ a width-based pruning threshold of \SI{0.15}{\milli\metre} combined with a width significance criterion that keeps only features with a width of \SIrange{0.19}{0.21}{\milli\metre}. The parameters differ only in the segmentation polarity (dale ``D'' or hill ``H''), the evaluated feature attribute, and the statistic aggregated across all grooves of a profile.

Two of the attributes, $R_\mathrm{dt}$ and $R_\mathrm{dq}$, are standardized in ISO~21920-2, while the remaining three are the custom groove-shape descriptors of Section~\ref{sec:single_param_search}. For a single mean-centred groove profile $z(x)$ sampled at $\Delta x = \SI{0.5}{\micro\metre}$, the attributes are defined as follows:
\begin{itemize}
\item $R_\mathrm{dt} = \max |z'(x)|$ and $R_\mathrm{dq} = \sqrt{\overline{z'(x)^2}}$: the standardized maximum absolute and root-mean-square profile gradient.
\item $R_\mathrm{dt,right}$: the same maximum absolute gradient, but evaluated only over the trailing \SI{0.03}{\milli\metre} of the groove.
\item \emph{Secant} gradient: $\max_x |z(x+\ell) - z(x)|/\ell$ over a baseline $\ell$ of \SI{5}{\percent} of the groove width, a finite-baseline counterpart to $R_\mathrm{dt}$.
\item \emph{Symmetry}: the maximum of the normalized cross-correlation between $z(x)$ and its mirror image, equal to $1$ for a perfectly symmetric groove.
\item \emph{Symmetry (folding)}: the root-mean-square deviation between the left half of the groove and the mirrored right half, normalized by the groove peak-to-valley height, equal to $0$ for a perfectly symmetric groove.
\item \emph{Height span}: the mean height over a \SI{25}{\micro\metre} window on the trailing flank, taken relative to the groove mean.
\end{itemize}
That every cross-target parameter reduces to a gradient, symmetry, or height measure of the same trailing flank is the structural reason for the high collinearity (VIF up to 37) reported in Section~\ref{sec:additional}.

\begin{table*}[t]
\centering
\caption{Parameters retained by stability selection for all three wear indicators, with their formal feature-characterization descriptors according to ISO~21920-2.}
\label{tab:stable_set}
\begin{tabular}{l l}
\hline
Feature-characterization descriptor & Attribute \\
\hline
\texttt{FC; D; Width 0.15 mm; HDw 0.19--0.21\,mm; Rdt; Mean} & Maximum absolute gradient $R_\mathrm{dt}$ \\
\texttt{FC; D; Width 0.15 mm; HDw 0.19--0.21\,mm; Rdt\_right; Sum} & Trailing-flank gradient $R_\mathrm{dt,right}$ \\
\texttt{FC; H; Width 0.15 mm; HDw 0.19--0.21\,mm; Rdq; Min} & RMS gradient $R_\mathrm{dq}$ \\
\texttt{FC; H; Width 0.15 mm; HDw 0.19--0.21\,mm; Secant; Max} & Secant gradient \\
\texttt{FC; D; Width 0.15 mm; HDw 0.19--0.21\,mm; H\_span; Sum} & Trailing-flank height span \\
\texttt{FC; D; Width 0.15 mm; HDw 0.19--0.21\,mm; Symmetry; Min} & Symmetry \\
\texttt{FC; D; Width 0.15 mm; HDw 0.19--0.21\,mm; SymmetryFolding; Sum} & Symmetry (folding) \\
\hline
\end{tabular}
\end{table*}

\clearpage
\addtocounter{page}{-1}
\end{document}